\documentclass[prb,twocolumn,showpacs,floatfix]{revtex4-1}
\usepackage{graphics,epsfig,amsfonts,amssymb,amsmath,soul}
\begin{document}

\title{Fast vortex wall motion in wide permalloy strips\\ from double switching of the vortex core}

\author{Virginia Est\'evez}
%\email{virginia.esteveznuno@aalto.fi}
\author{Lasse Laurson}
%\email{lasse.laurson@aalto.fi}

\affiliation{COMP Centre of Excellence and Helsinki Institute of Physics,
Department of Applied Physics, Aalto University, P.O.Box 11100, 
FI-00076 Aalto, Espoo, Finland.}

\begin{abstract}
We study vortex domain wall dynamics in wide permalloy strips driven by applied magnetic
fields and spin-polarized electric currents. As recently reported [V. Est\'evez and 
L. Laurson, Phys. Rev. B, {\bf 93}, 064403 (2016)], for sufficiently wide strips and 
above a threshold field, periodic dynamics of the vortex core are localized in the vicinity 
of one of the strip edges, and the velocity drop typically observed for narrow strips 
is replaced by a high-velocity plateau. Here, we analyze this behavior in more detail 
by means of micromagnetic simulations. We show that the high-velocity plateau originates 
from a repeated double switching of the magnetic vortex core, underlying the periodic 
vortex core dynamics in the vicinity of the strip edge, i.e., the ``attraction-repulsion'' 
effect. We also discuss the corresponding dynamics driven by spin-polarized currents, as 
well as the effect of including quenched random structural disorder to the system.
\end{abstract}
\maketitle

\section{Introduction}

In the last decade, domain wall (DW) dynamics in ferromagnetic nanostructures have 
received considerable attention, especially due to technological applications 
in spintronics, including memory\cite{parkinracetrack,barnes2006} and logic 
devices\cite{cowburn2000,allwood2002,allwood2005}.  Typically, DW dynamics in 
nanostrips and wires is induced by external magnetic fields\cite{atkinson2003,beach2005,nakatani2005,
reviewthiaville,hayashi2007,yang2008,moriya2010, ZIN-11,hayashi2008,nakatani2003,nosotrosdynamics} 
or spin-polarized electric currents\cite{spcklaui2003,thiaville2004,klaui2005,
vernier2004,thiaville2005,yamaguchi2004}. Details of the resulting DW dynamics 
depend on the DW structure, which in turn depends on the material and geometry being
considered. For materials with a high perpendicular magnetic anisotropy\cite{buschowbook,
coeybook}, the typical DW structures are of the Bloch 
and/or N\'eel type. On the other hand, for soft magnetic materials such as permalloy, 
the negligible magnetocrystalline anisotropy implies that the physics of the system 
is dictated by the competition between exchange interactions and shape anisotropy due to 
magnetostatic effects. The resulting in-plane domains along the long axis of the strip 
are separated by DWs with equilibrium structures ranging from relatively simple 
transverse and vortex DWs to more complex multi-vortex walls when increasing the strip 
width from tens and hundreds of nanometers towards several micrometers \cite{nakatani2005,
reviewthiaville,McMichael97,klaui2004,nosotrosequilibrium}. 

Most of the previous works related to in-plane materials have focused on the study 
of DW dynamics in narrow permalloy strips where the equilibrium DW structure is the 
transverse or vortex wall\cite{nakatani2005,reviewthiaville,beach2005,hayashi2007,yang2008,
moriya2010,lee2007,weerts2008,tretiakov2008, clarke2008}. The DW 
dynamics for these systems is characterized by the emergence of an instability, often referred 
to as the Walker breakdown\cite{walker74}, which appears when the DW internal degrees of 
freedom are excited by a strong enough external driving force. The Walker breakdown occurs for 
both magnetic fields $B_\text{ext}$ above the Walker field $B_\text{W}$ and spin-polarized 
current densities $J$ exceeding the Walker current density $J_\text{W}$\cite{beach2005,
hayashi2007,glathe2008,hayashi2008}. For low magnitudes of the driving force, i.e., within 
the steady or viscous regime, the DW propagation velocity increases monotonically up to 
$B_\text{ext}=B_\text{W}$ or $J=J_\text{W}$. Above the Walker breakdown, DW velocity drops 
abruptly as a consequence of the periodic transitions between different DW structures. 
For narrow strips with transverse DWs, repeated transitions between the transverse DW and 
a dynamic structure with an antivortex propagating across the strip width 
occur\cite{reviewthiaville}. For slightly wider strips (typically in the range of 200 to 300 
nm) with vortex wall as the equilibrium DW structure, similar periodic transitions between 
different DW structures including vortex, transverse and antivortex DWs have been 
observed\cite{reviewthiaville,lee2007}.      

Recently, we have shown that DW dynamics in even wider permalloy strips, with strip widths 
ranging from a few hundred nanometers to a few micrometers, display a multitude of
regimes, depending on the geometry of the system, as well as on the magnitude of the 
applied magnetic field\cite{nosotrosdynamics}. These include, e.g., transitions between various 
DW structures, and the fact that for some combinations of the strip geometry and the applied 
field the system cannot support a compact DW. The dynamics of multi-vortex DWs (double and 
triple vortex DWs\cite{nosotrosdynamics}) is controlled by the polarities of the vortices. 
One of the most interesting behaviors characteristic of wide enough strips is an oscillatory 
motion of the vortex core close to one of the strip edges, which we refer to as the 
{\it attraction-repulsion effect}\cite{nosotrosdynamics}. There, the vortex core first approaches
the strip edge driven by the gyrotropic force, followed by a reversal of propagation direction
as the core is pushed back towards the middle of the strip. Then, the core starts another
approach towards the same strip edge, and the process repeats itself. Due to this effect the 
velocity drop typically associated with a Walker breakdown is absent in wide strips, leading 
instead to an extended high-velocity plateau in the velocity-field curve\cite{nosotrosdynamics}. 
Somewhat similar absence of the velocity drop has been previously observed in systems with a 
high perpendicular anisotropy and a finite Dzyaloshinskii-Moriya interaction \cite{thiaville_dmi},
as well as in ferromagnetic nanotubes \cite{nanotube_apl}. However, many details of the 
attraction-repulsion effect are not fully understood, and thus more studies are required.

In this paper, we provide a complete description of the attraction-repulsion effect of vortex
DW dynamics by micromagnetic simulations. We analyze in detail the behavior when the DW dynamics 
is induced by an external magnetic field, and also consider the corresponding DW dynamics
driven by spin-polarized electric currents. We also address the question of how the presence of
quenched structural random disorder, originating, e.g., from the polycrystalline structure of
the strips, affects the behavior. We find that the attraction-repulsion effect arises due to 
repeated double switching of the magnetic vortex core in the vicinity of one of the strip 
edges, both for field and current drive. Although structural disorder affects the DW dynamics in 
these systems, the attraction-repulsion effect is stable against structural disorder of moderate 
strength.   

The paper is organized as follows: The following section (Section \ref{micromagn}) describes 
the system under study and the main features of our micromagnetic simulations. Section 
\ref{field} discusses in detail the attraction-repulsion effect when the DW dynamics is 
driven by an external magnetic field. DW dynamics driven by spin-polarized electric current 
in wide permalloy strips is described in Section \ref{current}, while Section \ref{disorder} is 
dedicated to the effect of structural disorder on the DW dynamics. The conclusions and a 
summary of the main results are given in Section \ref{summary}.

\begin{figure}[t!]
\leavevmode
\includegraphics[trim=1cm 0cm 1cm 2cm,clip,width=0.9\columnwidth]{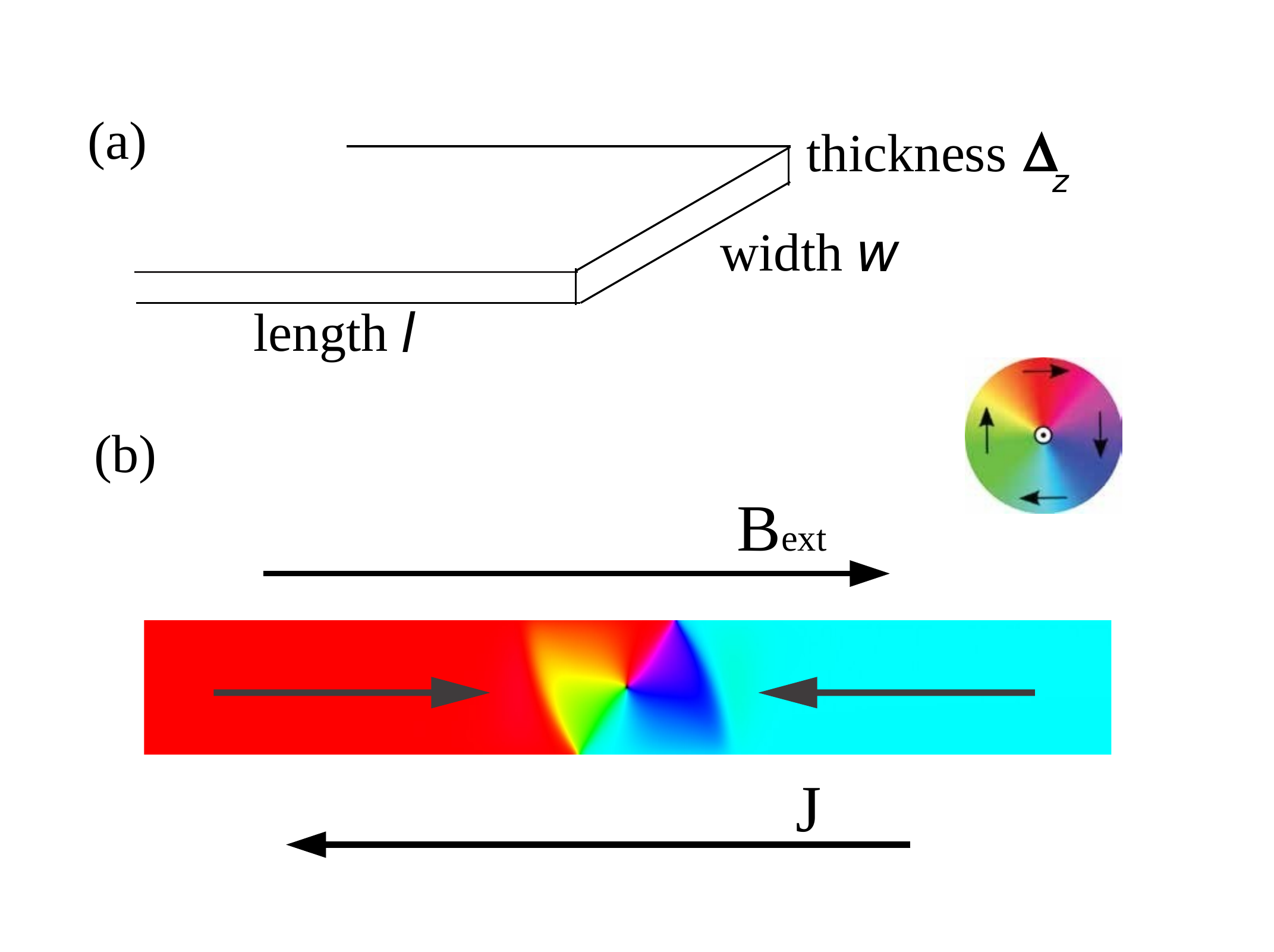}
\caption{(color online) (a) Geometry of the permalloy strip. 
(b) A top view of the magnetization in the initial state. Magnetization points 
along the long axis of the strip within the two domains (as indicated by the 
arrows) forming a head-to-head configuration. These domains are separated by a vortex 
DW. To study the DW dynamics, an external drive, i.e., either a magnetic field 
$B_\text{ext}$ or a spin-polarized electric current density $J$ is applied along the 
long axis of the strip.}
\label{fig:Fig1_paperattrepul}
\end{figure}

\section{Micromagnetic simulations}
\label{micromagn}

The system analyzed is a permalloy strip of width $w$ and thickness $\Delta_z$, satisfying
$\Delta_z \ll w$, see Fig.~\ref{fig:Fig1_paperattrepul} (a). DW dynamics is studied by micromagnetic
simulations, where in order to mimic an infinitely long strip, the magnetic charges are 
compensated on the left and right ends of the strip. We consider the length $l$ of the strip depending 
on the simulation window. For systems analyzed with a window moving with and centered at the DW, 
the length satisfies $l\geq 16w$. For a fixed simulation window we consider $l \geq 60w$.  The initial 
state is an in-plane head-to-head equilibrium DW structure separating the two head-to-head domains,
see Fig.~\ref{fig:Fig1_paperattrepul} (b). The equilibrium DW structure in general 
depends on $w$ and $\Delta_z$\cite{reviewthiaville,nosotrosequilibrium}, and is the DW structure 
with the lowest energy for a given system. Here, the vortex wall is the equilibrium DW structure 
for all $w$ and $\Delta_z$ values considered\cite{nosotrosequilibrium}. Starting from such an equilibrium state, DW
dynamics is induced by applying either an external magnetic field $B_\text{ext}$ or a spin-polarized
current density $J$, see Fig.~\ref{fig:Fig1_paperattrepul} (b). All the results presented in 
this work have been calculated for the typical material parameters of permalloy, i.e., 
saturation magnetization $M_\text{s}=860 \times 10^3$ A/m, exchange constant $A_\text{ex} = 13 
\times 10^{-12}$ J/m and the Gilbert damping constant $\alpha=0.01$. We analyze two different 
systems, the ideal case of strips free of any structural disorder or impurities, and strips 
with structural disorder. In the simulations the disorder is introduced by Voronoi 
tessellation\cite{nakatani2003,leliaert2014prb,leliaert2014,leliaert2014jap} to mimic
effects due to the polycrystalline structure of the strip, as is explained in more detail in 
Section \ref{disorder}. For all simulations we set the temperature $T$ equal to zero.

The simulations are performed using the GPU-accelerated micromagnetic code 
MuMax3 \cite{mumax3,mumax2011,mumax2014}, offering a significant speedup
as compared to CPU codes for the large system sizes we consider here.
The magnetization dynamics of the system is calculated numerically by the Landau-Lifshitz-Gilbert 
(LLG) equation \cite{gilbert2004,brownmicromagnetism},
\begin{equation}
\partial {\bf m}/\partial t =
\gamma {\bf H_{eff}} \times {\bf m} + \alpha {\bf m} \times
\partial {\bf m}/\partial t,
\end{equation}
 where ${\bf m}$ is the magnetization, $\gamma$ the 
gyromagnetic ratio, and ${\bf H}_\text{eff}$ the effective
field, with contributions due to exchange, Zeeman, and demagnetizing
energies. The discretization cell used depends on the system 
size, but is always bounded from above by the exchange length, $\Lambda = (2A/\mu_0 M_s^2)^{1/2}
\approx 5$ nm, in the in-plane directions, and equals $\Delta_z$ in the the out-of-plane
direction. 

\begin{figure}[t!]
\leavevmode
\includegraphics[clip,width=0.48\textwidth]{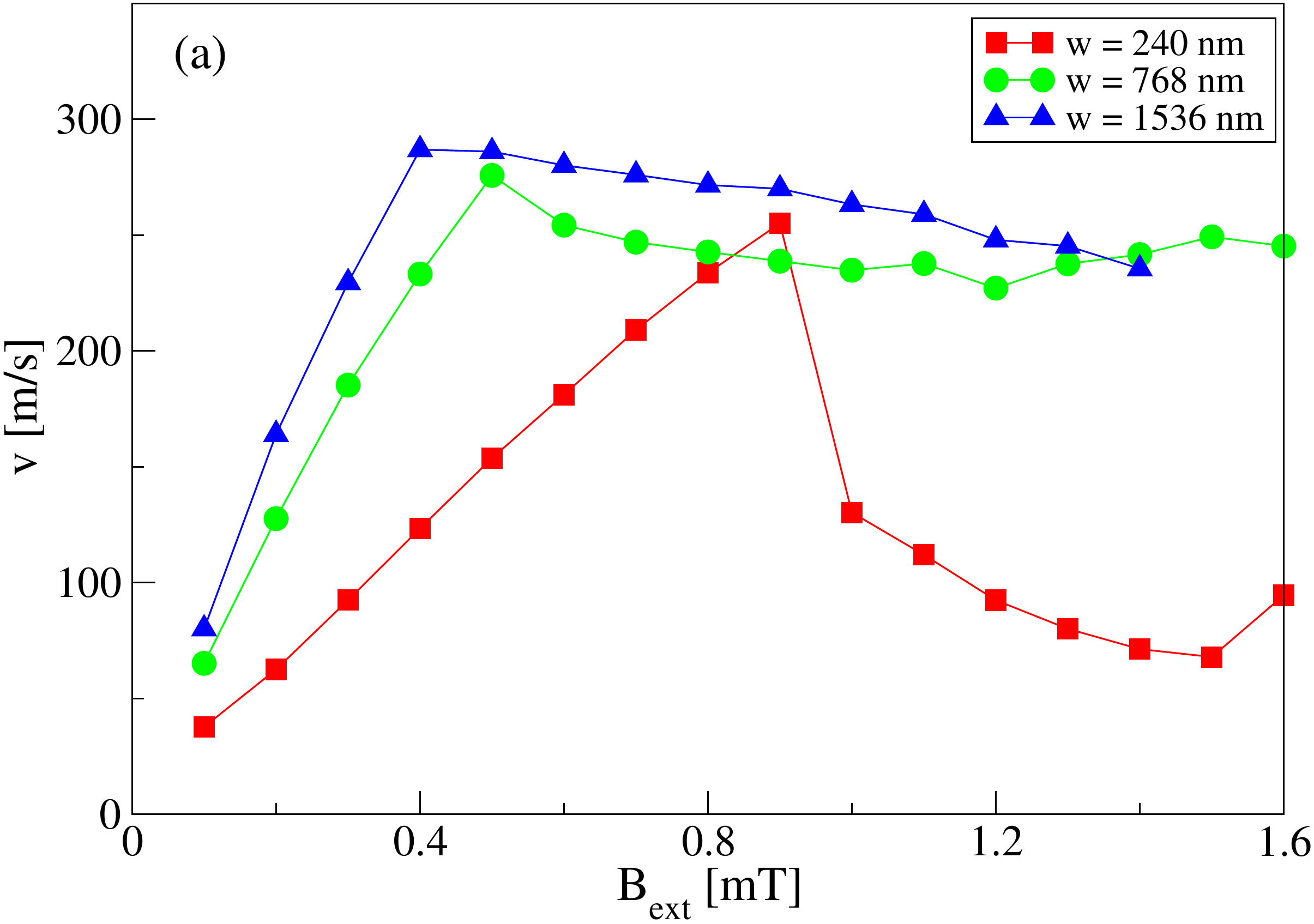}
\includegraphics[clip,width=0.48\textwidth]{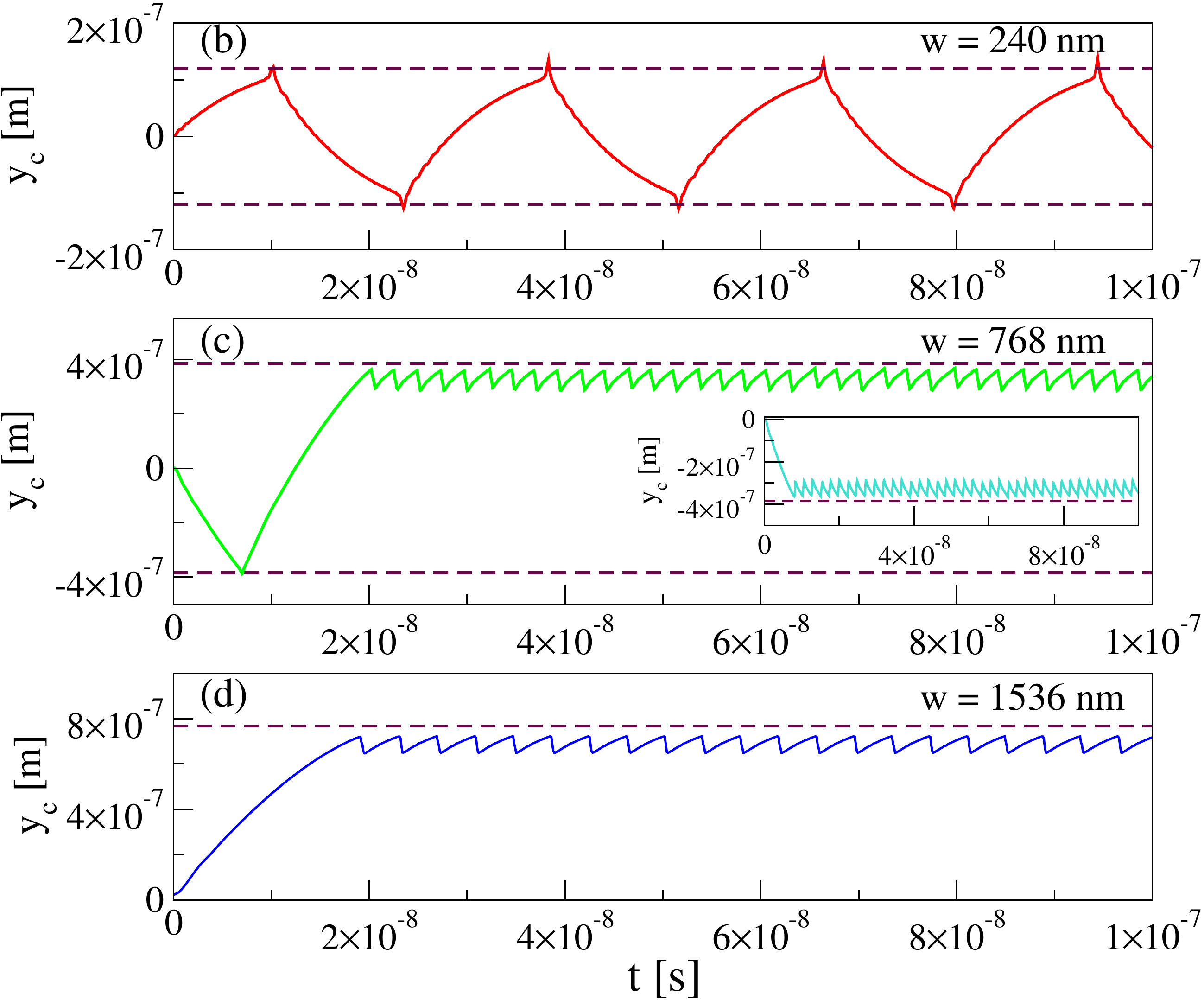}
\caption{(color online) (a) $v(B_\text{ext})$ curves for different strip geometries, 
where the vortex wall is the equilibrium DW structure. (b)-(d) The trajectory $y_\text{c}$ of 
the vortex core for the systems shown in (a) at a given external field. The dashed lines 
represent the strip edges. (b) $w = 240$ nm, $\Delta_z = 20$ nm and $B_\text{ext} =1.5$ mT. 
The vortex core moves repeatedly across the strip from one edge to the other. In wider strips,
the attraction-repulsion effect takes place:
(c) $w = 768$ nm, $\Delta_z = 15$ nm and $B_\text{ext} =1.5$ mT. The main panel shows the trajectory
of the core of a clockwise vortex, while the inset displays the corresponding $y_\text{c}(t)$ for the
core of a counter-clockwise vortex. (d) $w = 1536$ nm, $\Delta_z = 15$ nm and $B_\text{ext} =0.7 $ mT. 
$y_\text{c} = 0$ corresponds to the initial equilibrium state with the vortex core in the middle
of the strip, see Fig.~\ref{fig:Fig1_paperattrepul}(b).}
\label{fig:Fig2_paperattrepul}
\end{figure}

\section{Magnetic field driven DW dynamics}
\label{field}

As the attraction-repulsion effect is a behavior observed in the vortex DW dynamics, 
we restrict the study to strips with geometries where the vortex DW is the equilibrium 
DW structure\cite{nosotrosequilibrium}. The vortex DW is characterized by the core polarity 
$p$, and chirality $C$, or the sense of in-plane rotation around the core. The polarity is 
given by the sign of the out-of-plane magnetization of the core, and can thus be up or down, 
corresponding to  $p = \pm 1$. $C$ is defined as clockwise ($C=-1$) or counter-clockwise 
($C=+1$) rotation of the in-plane magnetization around the core. A clockwise VW with 
$C=-1$ and $p=-1$ is represented in Fig.~\ref{fig:Fig1_paperattrepul} (b).

For reference, we first review the typical dynamical behavior for field-driven vortex DW
dynamics in narrow permalloy strips. Fig.~\ref{fig:Fig2_paperattrepul} (a) shows the DW 
velocity $v$ as a function of the applied magnetic field $B_\text{ext}$ for three different 
strip geometries. In the case of a narrow strip of width $w=240$ nm and thickness 
$\Delta_z=20$ nm, a steady or viscous low-field regime with $v$ increasing 
roughly linearly with $B_\text{ext}$ up to the Walker field $B_\text{W}$ is observed. 
In this regime, the motion of the vortex DW occurs with the vortex core displaced 
from its equilibrium position in the middle of the strip, with the displacement increasing
with $B_\text{ext}$. For $B_\text{ext}>B_\text{W}$, $v$ drops abruptly as a consequence 
of the onset of periodic transformations between different DW structures. For example, 
for $B_\text{ext}=1.5$ mT, the vortex core repeatedly moves across the strip width from one 
edge to the other, leading to periodic transformations between vortex and transverse DW 
structures. The corresponding trajectory ($y$-coordinate) $y_\text{c}$ of the vortex core 
is shown in Fig.~\ref{fig:Fig2_paperattrepul} (b). The vortex core first approaches one of
the strip edges, and exits the strip when reaching the edge. It is subsequently injected 
back into the system with a reversed polarity, after which it moves towards the other edge, 
and the process repeats itself.

For wider strips, the $v(B_\text{ext})$ curve is very different to the one found for narrow 
systems, see Fig.~\ref{fig:Fig2_paperattrepul} (a) which also shows the $v(B_\text{ext})$ curves 
for two different wider strips. These curves have been reproduced here for reference from
our recent paper\cite{nosotrosdynamics}. The two wide strips considered have the same thickness 
$\Delta_z=15$ nm and different widths, $w=768$ nm and $w=1536$ nm. As in the case of 
narrow strips, for low fields $v$ displays a linear dependence on $B_\text{ext}$ within the 
steady or viscous regime. However, above this regime, a high-velocity plateau is observed
in the $v(B_\text{ext})$ curve, instead of the velocity drop found for narrow strips. 
A similar plateau was previously obtained 
experimentally and numerically\cite{weerts2008,ZIN-11}. The existence of this plateau was
associated with the fact that in this regime the vortex cannot leave the system through 
one of the edges of the strip\cite{ZIN-11}. However, we have recently shown that the 
reason for these large $v$-values is the attraction-repulsion effect\cite{nosotrosdynamics}. 
This behavior avoids the periodic transformations between different DW structures that 
reduce the DW propagation velocity above the Walker field $H_\text{W}$ in narrow strips. 
Moreover, for wide strips the vortex core is not able to leave the strip. For some reason, 
the vortex core is pushed back a small distance inside the strip before reaching the strip edge, and 
subsequently starts a new approach towards the same edge, before being pushed back again, 
(see also movie 1 provided as Supplemental Material\cite{SM}). 
Thus, the vortex core exhibits periodic oscillations confined within the vicinity of
one of the edges of the strip. As a result, the dynamics involving transformations 
between different DW structures found in narrow strips do not occur, and the velocity 
drop is absent, see Fig.~\ref{fig:Fig2_paperattrepul} (c)-(d).

A closer look reveals further details of this process. From the trajectory $y_\text{c}$ 
represented in the main panel of Fig.~\ref{fig:Fig2_paperattrepul} (c) it is evident that 
the vortex core first moves to the bottom edge, leaves the strip without any problem
(i.e., no attraction-repulsion effect takes place), is injected back into the system 
(with a reversed polarity), moves towards the top edge, and experiences the 
attraction-repulsion effect there. Thus, the attraction-repulsion effect only occurs in 
one of the edges of the strip. We find that the edge where this effect occurs depends on 
the vortex chirality $C$ and the direction of the applied magnetic field (positive or 
negative $x$ direction). For the same strip geometry and applied field, the vortex oscillations
of the attraction-repulsion effect take place close to the top or bottom edge depending on 
the chirality of the vortex. In Fig.~\ref{fig:Fig2_paperattrepul} (c), the main panel 
corresponds to a clockwise vortex ($C=-1$), whereas the inset shows the trajectory of a 
counter-clockwise ($C=+1$) vortex.

\begin{figure}[t!]
\leavevmode
\includegraphics[trim=0cm 0cm 0cm 0.5cm,clip=true,width=0.5\textwidth]{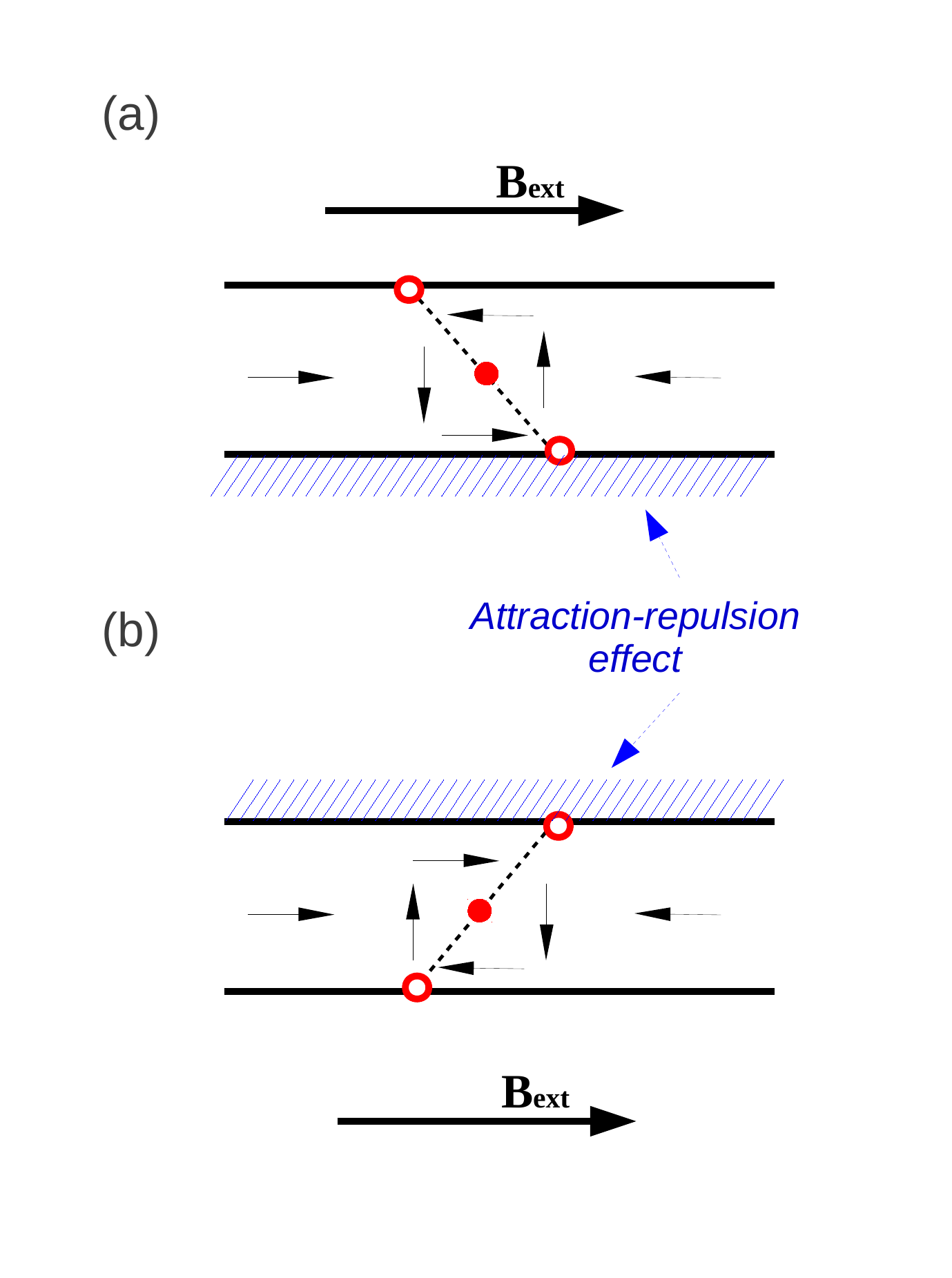}
\caption{(color online) Top view sketch of the vortex DW structures, indicating the strip edge
where the  attraction-repulsion effect takes place depending on the chirality of the vortex. 
(a) Counter-clockwise chirality. (b) Clockwise chirality. Vortex core and the half-antivortex 
edge defects are denoted by filled and open circles, respectively.}
\label{fig:Fig3_paperattrepul}
\end{figure}

To understand this, we start by considering the counter-clockwise vortex represented 
schematically in Fig.~\ref{fig:Fig3_paperattrepul} (a). Similar ideas to explain the 
fact that the vortex core cannot leave the system through one of the strip edge have
been presented by Zinoni {\it et al.}\cite{ZIN-11}. In Fig.~\ref{fig:Fig3_paperattrepul}, 
a full circle represents the vortex core, and the empty ones denote the half-antivortex edge defects. 
When the field is applied, the vortex core moves in the trajectory between the two edge defects, 
represented by the dashed line. If the vortex goes to the bottom edge, the area of 
the vortex structure with magnetization parallel to the applied field decreases, leading 
to a cost in Zeeman energy that prevents the exit of the core\cite{ZIN-11}. Instead, 
when the vortex moves towards the top edge the region with the vortex magnetization 
antiparallel to the field shrinks, while the parallel area increases, facilitating the 
core expulsion off the strip. Analogously, for a clockwise vortex [see 
Fig.~\ref{fig:Fig3_paperattrepul}(b)], the area of the vortex magnetization 
parallel to $B_\text{ext}$ decreases when the vortex moves towards the top edge. Thus, 
the edge in the neighborhood of  which the attraction-repulsion effect occurs is 
different depending on the vortex chirality. If the magnetic field is applied in the 
opposite sense, the strip edge associated with the attraction-repulsion effect changes 
for a given chirality. One should note that the relation between the chirality and 
the field is the same in narrow and wide strips. However, the significant energy cost 
preventing the expulsion of the vortex is only characteristic of wide enough strips. 
This is due to shape anisotropy (understood here as the tendency of the magnetization to
align along the long axis of the strip) being essentially an edge effect.
While in narrow strips the shape anisotropy (or the magnetostatic energy) is more uniform throughout 
the whole strip, in wide strips shape anisotropy favoring magnetization to point along 
the strip axis is less important in the middle of the strip than close to the edges. 
As a result, in wide strips the energy required to push the vortex core out of the strip 
is larger than the energy needed to move the vortex core off its equilibrium position 
in the middle of the strip. 

This argument explains why the vortex core cannot leave the strip. However, as we can 
see in Fig.~\ref{fig:Fig2_paperattrepul}(c)-(d), the oscillations occur in the vicinity 
of the strip edge, and in no case the vortex core arrives to the opposite edge. Thus, 
something is happening during the motion of the vortex core towards the opposite edge,
leading to a reversal in the direction of motion. Moreover, for the case of a strip 
with $w=768$ nm and $\Delta_z=15$ nm, for some values of the field the attraction-repulsion 
effect occurs even if the vortex core leaves the strip for a short time. Thus, a 
pertinent question is: why does the core reverse its propagation direction back towards
the strip edge?

\begin{figure}[t!]
\leavevmode
\includegraphics[clip,width=0.5\textwidth]{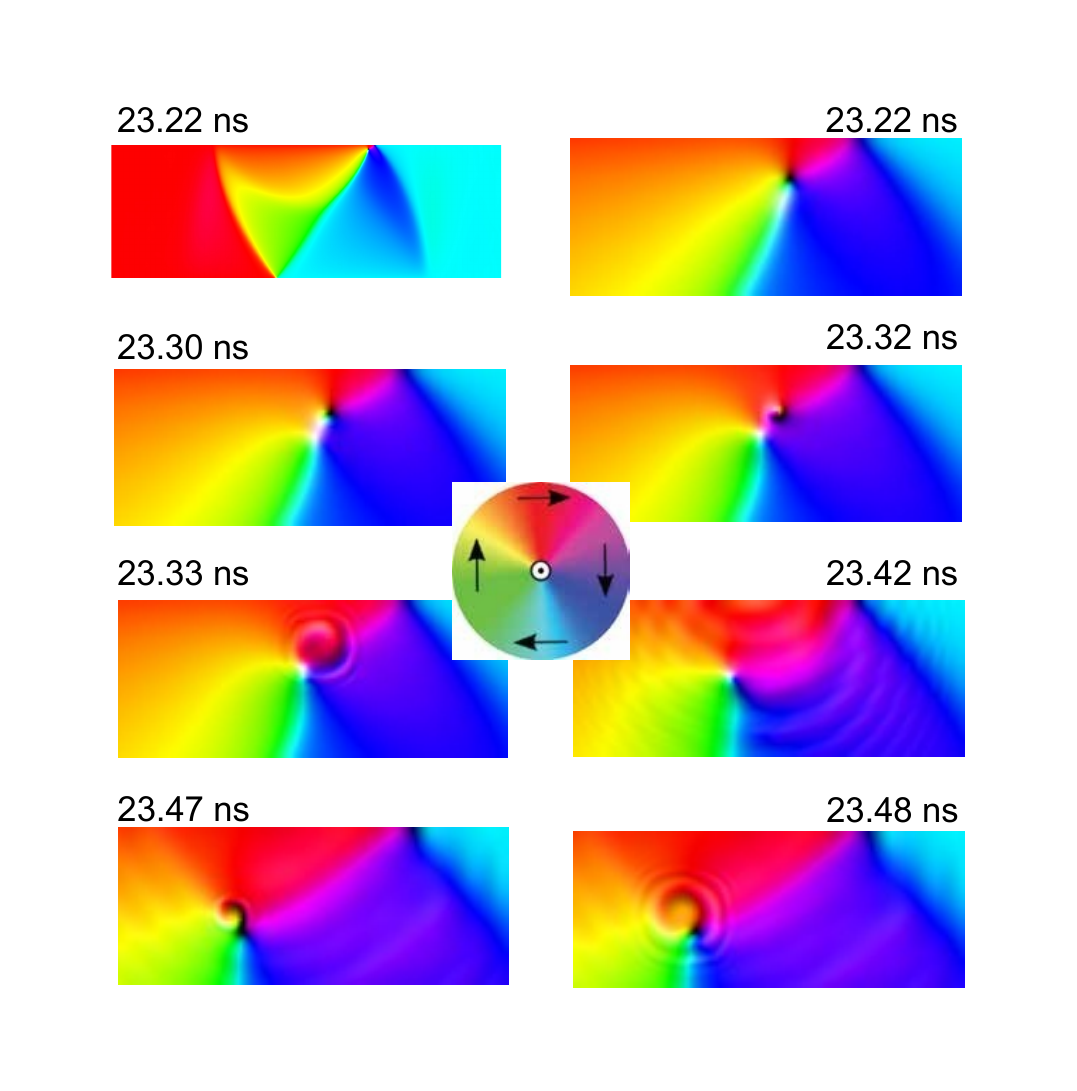}
\caption{(color online) A sequence of magnetization snapshots with a high temporal 
resolution showing the different stages of the fast double switching event underlying the 
attraction-repulsion effect. Here, $B_\text{ext} = 0.7$ mT in a strip with $w =1536$ nm 
and $\Delta_z = 15$ nm is considered. The top-left panel shows the whole vortex wall,
while the other panels are magnifications of the region around the vortex core close to
the top edge of the strip. The process consists of a formation of a dip particle, leading
to the nucleation of a vortex-antivortex pair, followed by annihilation of the initial
vortex and the antivortex (first switching). The second core switching is assisted by the
spin waves generated in the annihilation process.}
\label{fig:Fig4_paperattrepul}
\end{figure}

When an external field is applied, the motion of the vortex core is controlled by 
the gyrotropic force\cite{malozemoff1979}  ${\bf F}^\text{g}=pG {\bf \widehat{z}} \times {\bf v}$; 
here $p$ is the polarity, $G$ is the gyrotropic constant and ${\bf v}$ the velocity of the 
vortex core. For a given field
along the long axis of the strip, depending on the polarity the vortex core moves towards 
the top or bottom edge of the strip. Thus, a change of direction in the motion of the vortex 
can only be due to a change in the vortex polarity. Analyzing the DW dynamics with more 
precision (high temporal resolution), we have observed that the attraction-repulsion effect 
involves a double switching of the magnetic vortex core. Fig.~\ref{fig:Fig4_paperattrepul} 
shows the fast VW dynamics sequence during such a process
for a strip with $w=1536$ nm and $\Delta_z=15$ nm for $B_\text{ext}=0.7$ mT, 
see also movie 2 provided as Supplemental Material\cite{SM}. The top left snapshot in 
Fig.~\ref{fig:Fig4_paperattrepul} shows a top view of the strip where the vortex core is 
very close to the top edge of the strip, driven towards it by the gyrotropic force. 
Magnifying the area close to the vortex and the strip edge (top right panel of 
Fig.~\ref{fig:Fig4_paperattrepul}) shows that close to the out-of-plane
vortex core (black) appears an area with opposite out-of-plane magnetization (white). 
This deformation of the core is known as a dip particle \cite{waeyenberge2006,guslienko2008}. With 
time this area increases and leads to the nucleation of a vortex-antivortex pair, with both the
new vortex and antivortex having the same polarities which is opposite to the polarity of 
the initial vortex ($23.30$ ns, Fig.~\ref{fig:Fig4_paperattrepul}). Next, a very fast 
annihilation process between the antivortex and the initial vortex occurs ($23.32$ ns, 
Fig.~\ref{fig:Fig4_paperattrepul}). Thus, the vortex remaining in the strip has the opposite 
polarity to the initial vortex, leading to a polarity switching of the vortex core.
Consequently, the gyrotropic force changes direction, and the vortex core starts to move
away from the strip edge. The vortex-antivortex annihilation process is followed by 
emission of spin waves\cite{lee2005,hertel2006,waeyenberge2006,tetriakov2007}, ($23.33$ ns, 
Fig.~\ref{fig:Fig4_paperattrepul}). This is due to violation of the conservation of
the skyrmion charge, $q=np/2$, where $n$ is the winding number\cite{TCH-05}, in the 
antiparallel annihilation process of the vortex-antivortex pair. 
As a consequence of these spin waves, very close to the core of the remaining vortex, an 
area with out-of-plane magnetization opposite to that of the vortex core is created 
($23.42$ ns, Fig.~\ref{fig:Fig4_paperattrepul}). This area grows, becoming larger than 
the vortex core. Soon after, the area wraps the vortex core by forming a spiral around it ($23.47$ ns, 
Fig.~\ref{fig:Fig4_paperattrepul}). This process not only produces more spin waves, but 
also leads to the second switching of the vortex core polarity ($23.48$ ns, 
Fig.~\ref{fig:Fig4_paperattrepul}). This second reversal of the polarity occurs without 
the formation of a vortex-antivortex pair. Given the polarity of the vortex, the 
gyrotropic force now pushes the core towards the strip edge, and the process repeats itself. 

We note that for some strip geometries and applied fields, for example $w=768$ nm, $\Delta_z=15$ nm 
and $B=1$ mT, a similar process is found where no dip particle is formed. In this case, 
the first core switching takes place when the vortex core is able to exit the 
strip and is subsequently injected back with a reversed polarity. At the same time, there is injection 
of spin waves into the strip. This is then followed by 
the second reversal as described above for the ``pure'' attraction-repulsion effect, 
again leading to oscillatory vortex core dynamics localized to the neighborhood of one 
of the edges of the strip.

\begin{figure}[t!]
\leavevmode
\includegraphics[clip,width=0.48\textwidth]{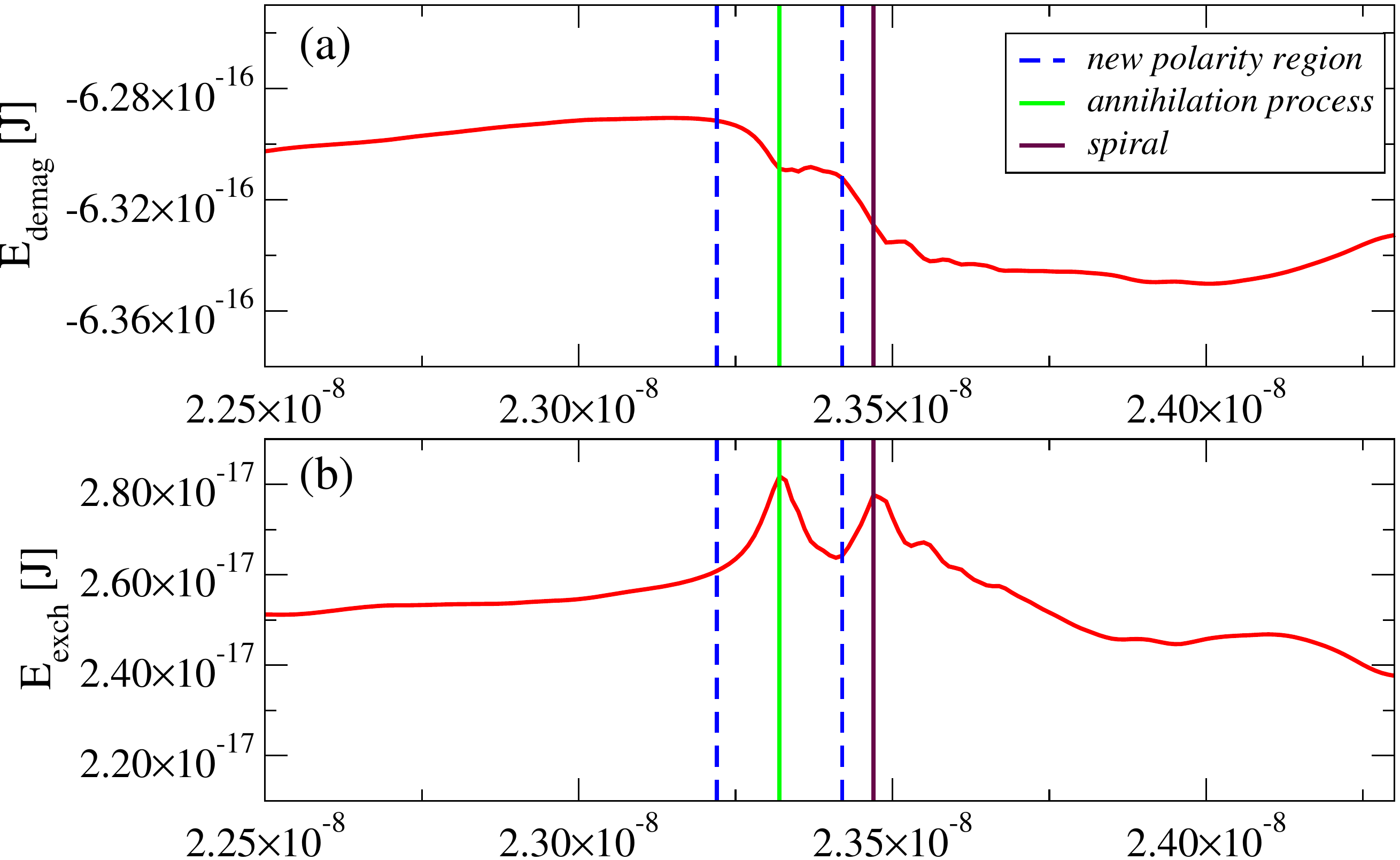}
\includegraphics[clip,width=0.48\textwidth]{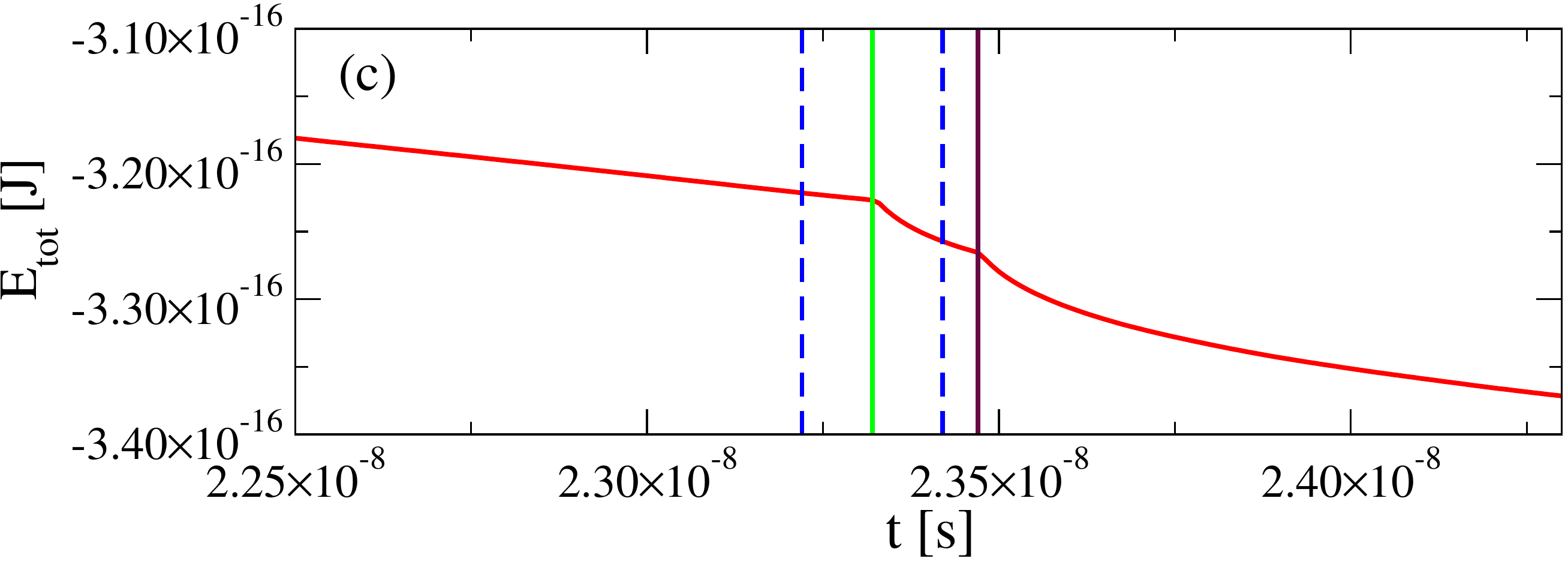}
\caption{(color online) Different energies of the system as a function of time during the double 
switching process portrayed in Fig. \ref{fig:Fig4_paperattrepul}, with $w =1536$ nm, $\Delta_z = 15$ nm 
and $B_\text{ext} = 0.7$ mT. (a) $E_\text{demag}$. (b) $E_\text{exch}$. 
(c) $E_\text{tot}=E_\text{demag}+E_\text{exch}+E_\text{Zeeman}$.}
\label{fig:Fig5_paperattrepul}
\end{figure}

Thus, the attraction-repulsion effect involves a double switching of the magnetic 
vortex core. The first switching proceeds via the creation of a vortex-antivortex pair 
followed by an annihilation process. The second one appears to be due to the spin 
waves generated in this annihilation process. The reversal processes occur very fast, 
with the double switching time being $\sim 260$ ps for the relatively low field considered
here. This time depends slightly on the geometry of the strip and the magnitude of 
the driving field. Due to its importance in 
magnetic data storage, switching of the magnetic vortex 
core has been studied extensively, and may be excited by magnetic fields\cite{waeyenberge2006,
vansteenkiste2009,hertel2007,guslienko2008,lee2007B,kim2008,xiao2006}, 
spin-polarized electric currents\cite{yamada2007,liu2007,yamada2008,yamada2010,
caputo2007,kim2007} and spin waves\cite{kammerer2011,kammerer2012,noske2016,
park2005,bauer2014, yoo2012,ruckriem2014}. The switching mechanism mediated by the 
vortex-antivortex pair is rather well-known. The creation of a dip particle leading to 
the nucleation of a vortex-antivortex pair has been associated with an effective field, 
the gyrofield, which depends on the vortex core velocity\cite{guslienko2008}. There is 
a widespread belief that the switching process occurs when the velocity of the vortex 
core reaches a critical ``core-switching'' velocity\cite{kim2007}. However, this is not 
true for all the cases. Several works\cite{yamada2010,kammerer2011,kravchuk2009} have 
shown that high velocity of the vortex core is not a necessary condition for core 
switching. Indeed, it has been shown that the vortex core reversal occurs when a
well-defined threshold in the exchange energy density is reached, corresponding to
the energy necessary for the production of a vortex-antivortex pair\cite{gigla2011}. 
In our system we have not observed a clear correlation between the velocity of the 
vortex core and its reversal, such that the velocity would increase prior to the switching 
event. In our case, the creation of the dip particle appears to be related to the fact 
that the vortex core cannot exit the strip due to increasing energy as the core approaches 
the strip edge. The core reversal would then serve as a mechanism to reduce the energy 
of the system.
  
To quantify this, Fig.~\ref{fig:Fig5_paperattrepul} shows the time evolution of demagnetizing energy 
$E_\text{demag}$, the exchange energy $E_\text{exch}$ and the total energy $E_\text{tot}$ 
(with $E_\text{tot} = E_\text{demag} + E_\text{exch} + E_\text{Zeeman}$) during a double 
switching process for a strip with $w=1536$ nm, $\Delta_z=15$ nm and $B_\text{ext}=0.7$ mT. 
Comparing the times with those in Fig.~\ref{fig:Fig4_paperattrepul}, we can see that as the 
vortex core is approaching the strip edge, both $E_\text{demag}$ and $E_\text{exch}$ increase. 
In addition, as discussed
above, also $E_\text{Zeeman}$ increases locally as the vortex core is approaching the edge of the strip
where the attraction-repulsion effect takes place (see again Fig. \ref{fig:Fig3_paperattrepul}). The
decreasing {\it global} $E_\text{Zeeman}$ dominating the time-evolution of $E_\text{tot}$ in 
Fig.~\ref{fig:Fig5_paperattrepul} is a result of growth of the domain parallel to $B_\text{ext}$
due to domain wall motion. It is known that the nucleation of 
vortex cores may decrease $E_\text{demag}$ \cite{xiao2006}. As can be observed 
in Fig.~\ref{fig:Fig5_paperattrepul}(a), when the dip particle is created, $E_\text{demag}$ 
decreases until the annihilation process takes place. After that, the time-dependence of the 
demagnetizing energy displays a plateau, and decreases again when the new out-of-plane area 
is created close to the vortex core by the emission of spin waves. Simultaneously, 
$E_\text{exch}$ increases when the areas with opposite polarity are created, and only 
decreases when the annihilation process and the second switching (spiral) occur, see 
Fig.~\ref{fig:Fig5_paperattrepul}(b). After the creation of the dip particle, $E_\text{exch}$ 
increases as a consequence of the increased magnetization gradients due to the different 
(anti)vortex cores present in the strip after the nucleation. Although $E_\text{exch}$ increases 
with the creation of areas with opposite polarity, after the double switching $E_\text{exch}$ 
decreases. Fig.~\ref{fig:Fig5_paperattrepul}(c) shows that the total energy $E_\text{tot}$ 
decreases faster during the switching processes. Notice also that $E_\text{demag}$ starts to 
increase again after the double switching process as the vortex core starts another approach 
towards the strip edge. Thus, in our system the formation of the dip 
particle, and the subsequent core reversal, is a mechanism of the system to decrease its energy.

\section{Spin-polarized current driven DW dynamics}
\label{current}

\begin{figure}[t!]
\leavevmode
\includegraphics[clip,width=0.48\textwidth]{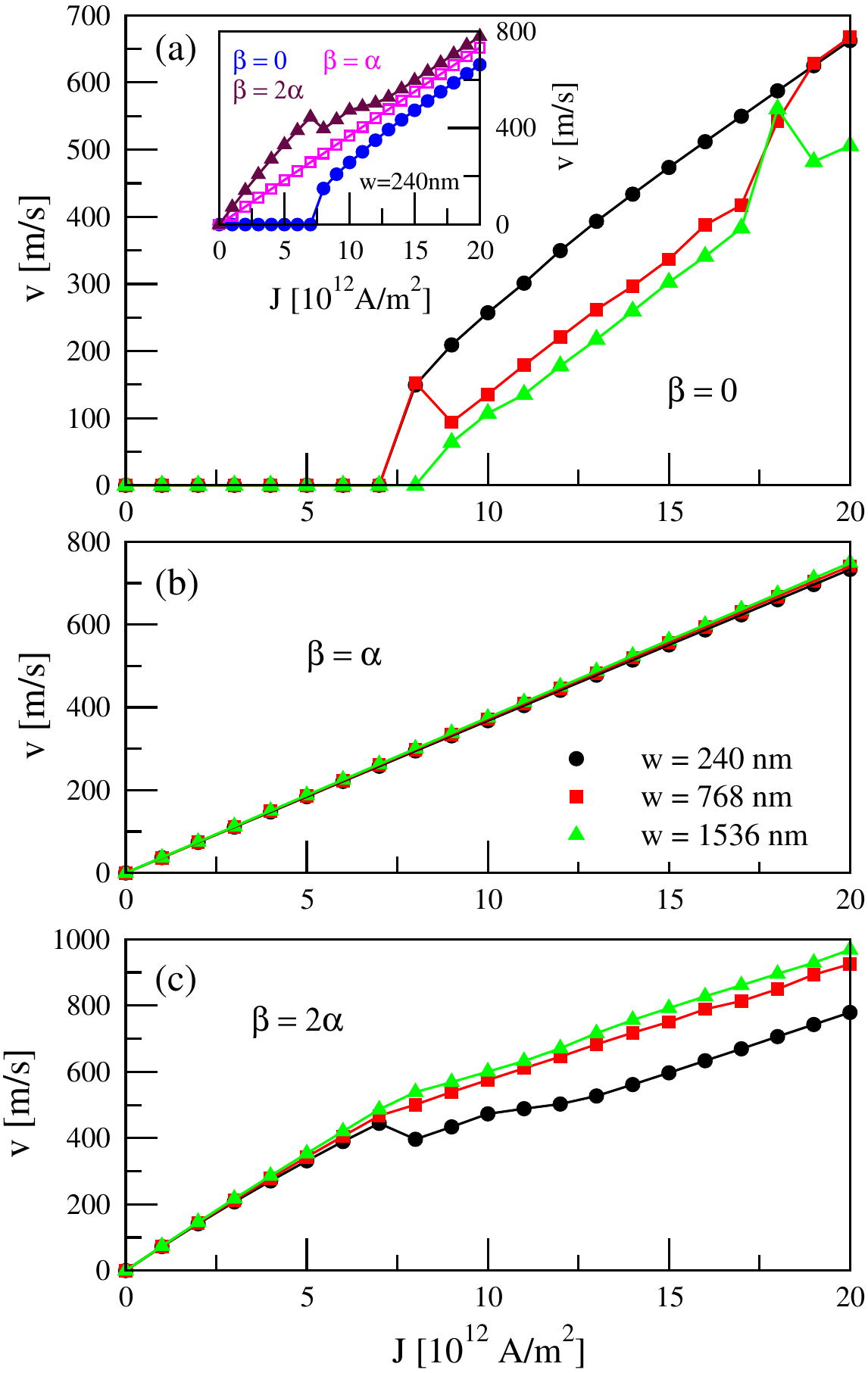}
\caption{(color online) Main figures: $v(J)$ curves for several values 
of $\beta$ and different strip geometries, where vortex wall is the equilibrium DW structure. 
The widths of the strips considered are $w = 240$ nm, $w = 768$ nm and $w = 1536$ nm. 
For $w =240$ nm the thickness is  $\Delta_z = 20$ nm, whereas for the other two strips 
$\Delta_z = 15$ nm. (a) $\beta = 0$. Inset: $v(J_\text{ext})$ for a strip with $w = 240$ nm, 
$\Delta_z = 20$ nm and several values of $\beta$. (b) $\beta = \alpha$. (c) $\beta = 2\alpha$.}
\label{fig:Fig6_paperattrepul}
\end{figure}
 
DW dynamics in conductive ferromagnetic strips can be driven also by spin-polarized electric 
currents\cite{spcklaui2003,vernier2004,klaui2005,thiaville2004,thiaville2005,yamaguchi2004}. 
So far, previous works have focused on narrow strips, and practically nothing is known 
about DW dynamics driven by spin-polarized electric currents in wide permalloy strips. 

In general, the nature of the spin-polarized current is quite different from a magnetic field, 
thus leading to differences in the resulting DW dynamics. To study the DW dynamics driven by 
spin-polarized electric currents, the LLG equation is modified to read\cite{thiaville2005}
\begin{eqnarray}
\partial {\bf m}/\partial t & = & 
\gamma {\bf H_{eff}} \times {\bf m} + \alpha {\bf m} \times
\partial {\bf m}/\partial t\\ \nonumber
& & -({\bf u}\cdot {\bf \nabla}){\bf m} + \beta{\bf m}\times [({\bf u} \cdot {\bf \nabla}){\bf m}],
\label{LLQspc}
\end{eqnarray}
where ${\bf u}$ is a vector in the direction of the electron flow given by 
${\bf u}=JPg\mu_B/(2eM_s)$. $e$ is the electric charge of the electron, $P$
is the polarization and $\beta$ a non-dimensional parameter describing the degree of 
non-adiabaticity of the spin-transfer torque. The nature of the DW dynamics depends 
strongly on $\beta$. Here, we analyze the DW dynamics for three cases, i.e., for $\beta=0$, 
$\beta=\alpha$ and $\beta=2\alpha$. For the polarization, we consider a typical value for 
the Permalloy, $P=0.56$. The direction of the current-induced DW motion is opposite to 
the current direction, and thus the spin-polarized current is applied along the 
long axis of the strip in the direction shown in Fig.~\ref{fig:Fig1_paperattrepul} (b). 

First, we again review the well-known DW dynamics in narrow strips for reference. Inset of 
Fig.~\ref{fig:Fig6_paperattrepul} (a) shows the DW velocity $v$ as a function of the 
current density $J$ for a strip with $w=240$ nm and $\Delta_z=20$ nm, with the vortex DW the 
equilibrium DW structure. For low current densities, i.e., below the intrinsic pinning
threshold \cite{zhang2004,thiaville2004}, the velocity is equal to zero in the adiabatic case 
($\beta=0$). Above this depinning threshold, $v$ increases with $J$, and for large enough $J$, 
exhibits a linear dependence on $J$. In this regime, the DW dynamics shows transformations 
between different DW structures, characteristic of the Walker breakdown. In the case of 
$\beta=2\alpha$, for low $J$, $v$ increases linearly up to the Walker current density 
$J_\text{W}$ similarly to the field-driven case. For both nonadiabatic cases ($\beta=\alpha$
and $\beta=2\alpha$), the velocity in this regime is given by $v=\beta u/\alpha$\cite{thiaville2005}. 
Above $J_\text{W}$, the velocity drops but less dramatically than in the field-driven system, 
see the inset of Fig.~\ref{fig:Fig6_paperattrepul} (a) and Fig.~\ref{fig:Fig2_paperattrepul} (a). 
Above the Walker breakdown $v$ increases roughly linearly with a different slope, with 
transformations between vortex and transverse DW structures taking place. The current 
density of the depinning threshold for $\beta=0$ and the Walker current density $J_\text{W}$ 
for $\beta=2\alpha$ coincide. For $\beta=\alpha$, a linear relation $v \approx \beta u/\alpha$ 
is obtained for any $J$, and the vortex DW moves with the vortex core in the middle of the strip.

\begin{figure}[t!]
\leavevmode
\includegraphics[clip,width=0.48\textwidth]{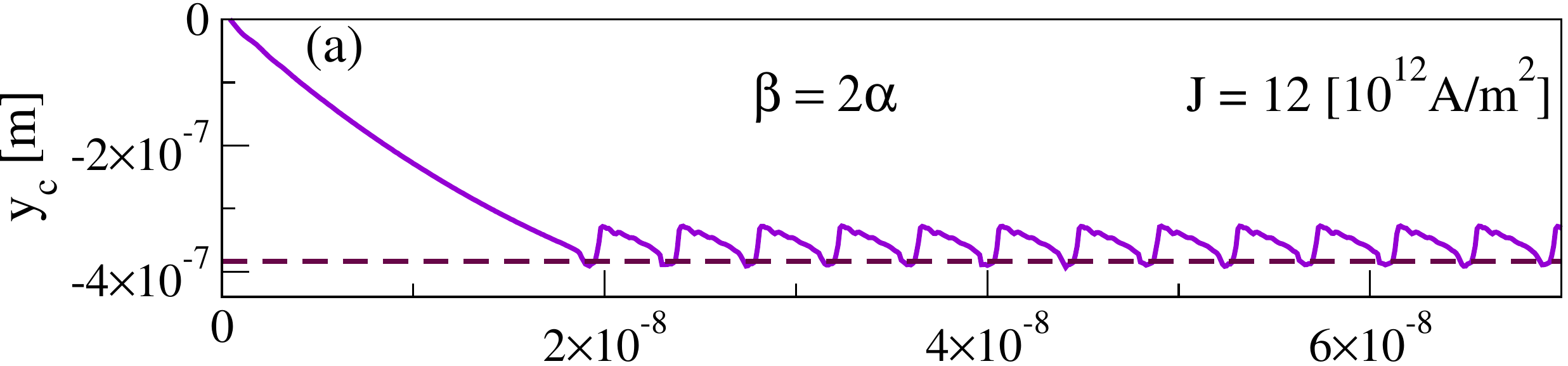}
\includegraphics[clip,width=0.48\textwidth]{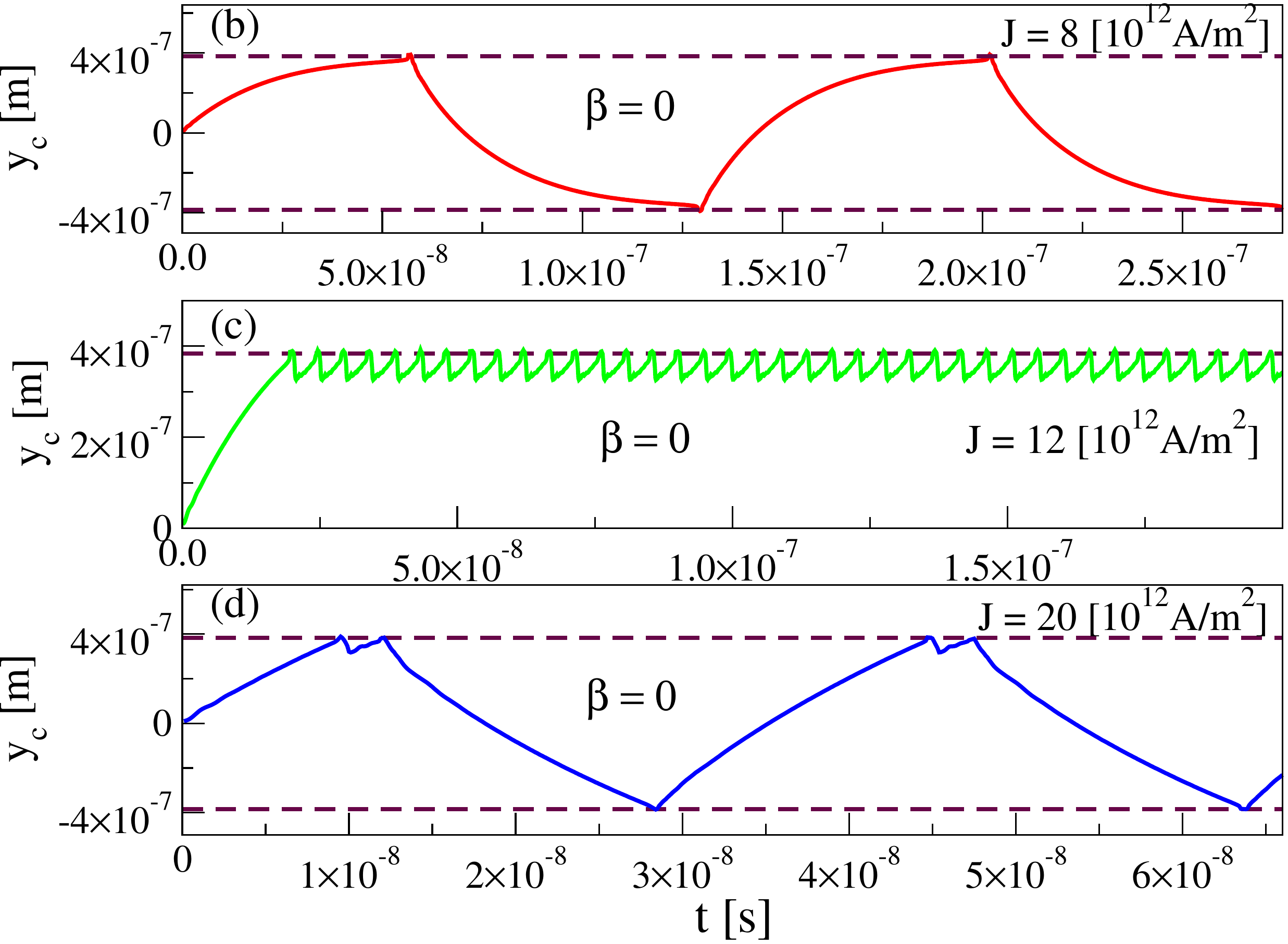}
\caption{(color online) Current-driven trajectories $y_\text{c}(t)$ of the vortex core for a strip with 
$w = 768$ nm and $\Delta_z = 15$ for different values of $J$ and $\beta$. The dashed lines 
represent the strip edges. (a) $\beta= 2 \alpha$ and $J = 12\times 10^{12}$ A/m$^{2}$, showing the 
attraction-repulsion effect. (b)-(d) Trajectories for different regimes of behavior in the case of 
$\beta=0$. (b) For $J = 8\times 10^{12}$ A/m$^{2}$, the vortex core moves repeatedly across the strip from one 
edge to the other, showing the typical behavior above the Walker breakdown. 
(c) For $J = 12\times 10^{12}$ A/m$^{2}$ the  attraction-repulsion effect is observed. 
(d) For $J = 20\times 10^{12}$ A/m$^{2}$, the trajectory displays a mixture between the 
transformations above the Walker breakdown and the attraction-repulsion effect.  
$y_\text{c} = 0$ corresponds to the initial equilibrium state with the vortex core in the middle
of the strip.}
\label{fig:Fig7_paperattrepul}
\end{figure}

Next, we proceed to consider wider permalloy strips. For $\beta=\alpha$, the DW 
dynamics is the same as that found for narrow strips. In fact, the behavior for $\beta=\alpha$ 
is fully independent of the width of the strip, as can be seen in 
Fig.~\ref{fig:Fig6_paperattrepul} (b), where the $v(J)$ curves are shown for different 
strip geometries. Notice that we consider here the same strips as in the field-driven case: 
$w=120$ nm and $\Delta_z=20$ nm, as well as $w=768$ nm and $w=1536$ nm both with $\Delta_z=15$ 
nm. Fig.~\ref{fig:Fig6_paperattrepul} (c) shows the $v(J)$ curves for $\beta=2\alpha$. For 
low values of $J$, the velocity shows a linear behavior independent of the 
strip width $w$. A key point here is that in wider strips the velocity does not 
exhibit a drop as in the narrow case. Instead, there is a change in the slope of the 
$v(J)$ curve around the current density corresponding to $J_\text{W}$ of the narrow system, 
see  Fig.~\ref{fig:Fig6_paperattrepul} (c). As in the field-driven case, the underlying
reason for this lack of velocity drop is the appearance of the attraction-repulsion effect. 
In the spin-polarized current driven case, the vortex core exits the strip more
frequently than in the field-driven case. Only for some current densities for the widest
strip ($w=1536$ nm), the ``pure'' attraction-repulsion effect is observed. This is probably due 
to the spin-polarized current acting directly on the vortex core (or, more generally, large 
magnetization gradients), while the magnetic field acts more uniformly on the whole DW 
structure. As a result, the vortex core can leave the strip more easily in the 
spin-polarized current driven case. The trajectory $y_\text{c}$ of the vortex core for a strip with 
$w=768$ nm and $\Delta_z=15$ nm is shown in Fig.~\ref{fig:Fig7_paperattrepul} (a) for
$J=12\times10^{12}$ A/m$^2$. The attraction-repulsion effect occurs at the bottom edge 
as the DW structure studied in this case is a counter-clockwise vortex DW. Depending on 
the value of $J$, the amplitude of the periodic oscillations can be smaller or larger. 
The larger the current density, the faster the double switching process.

For the adiabatic ($\beta=0$) case considered in Fig.~\ref{fig:Fig6_paperattrepul} (a), 
different regimes of behavior occur above the depinning threshold when the width of 
the strip increases. The depinning threshold appears to increase slightly with the strip width. 
For strips with $w=240$ nm and $w=768$ nm, the depinning threshold is found to be the same, 
whereas for $w=1536$ nm it is a little larger. For $J=8\times10^{12}$ A/m$^2$ in the case of a strip with $w=768$ nm, 
repeated transformations between vortex and transverse DW structures take place as in the case of narrow strips. The vortex core
dynamics (the core trajectory $y_\text{c}$) in the former case is shown in 
Fig.~\ref{fig:Fig7_paperattrepul} (b). 
For $8\times 10^{12} A/m^{2} \leq J \leq 16\times 10^{12} A/m^{2}$,
the vortex core moves to the strip edge and there is a kind of attraction-repulsion 
effect, with sometimes also a nucleation of an antivortex taking place. Contrary to the field-driven or the $\beta=2\alpha$ case, when the DW dynamics presents this kind of 
attraction-repulsion effect $v$ decreases with respect to the case in which there are 
transformations between DW structures. The reduction of the velocity is due to the nucleation of an antivortex during the attraction-repulsion effect. Moreover, in this regime the velocity decreases 
with the strip width. For $\beta=0$, when a spin-polarized current is applied, the vortex core 
moves in the opposite direction as compared to that of the field-driven or $\beta=2\alpha$ case. 
Figs.~\ref{fig:Fig7_paperattrepul} (a) and (c) show the $y_\text{c}(t)$ for the same strip for 
$J=12\times10^{12}$ A/m$^2$ and $\beta=2\alpha$ and $\beta=0$, respectively. The same DW structure 
has been considered in both cases, i.e.,  a vortex DW with the same chirality and polarity. As 
can be seen, the vortex core motion in both cases is opposite even if the polarity is the same, 
and the attraction-repulsion effect occurs in different strip edges. Considering LLG equation with 
the spin transfer torque terms\cite{zhang2004}, the fourth term gives an opposite 
sign for the adiabatic case with respect to the nonadiabatic cases with $\beta>\alpha$. This is the 
reason why in the adiabatic ($\beta = 0$) case the vortex core moves to the opposite direction to 
that of the $\beta=2\alpha$ case.
For $J\geq 18\times10^{12}$ A/m$^2$ in the case of $w=768$ nm, $v$ experiences a large increase, see 
Fig.~\ref{fig:Fig6_paperattrepul} (a). In this regime the DW 
dynamics shows a combination of the transformations between DW structures and the 
attraction-repulsion effect, see Fig.~\ref{fig:Fig7_paperattrepul} (d) where the corresponding 
$y_\text{c}(t)$ curve is shown for $J=20\times 10^{12}$ A/m$^2$. The attraction-repulsion effect is observed also for $w=1536$ nm. The 
combination of the attraction-repulsion effect and the transformations 
between DW structures only appears for $J=18\times10^{12}$ A/m$^2$. This again illustrates the idea
that as the strip width increases, the attraction-repulsion effect becomes increasingly prominent.

\begin{figure}[t!]
\leavevmode
\includegraphics[clip,width=0.48\textwidth]{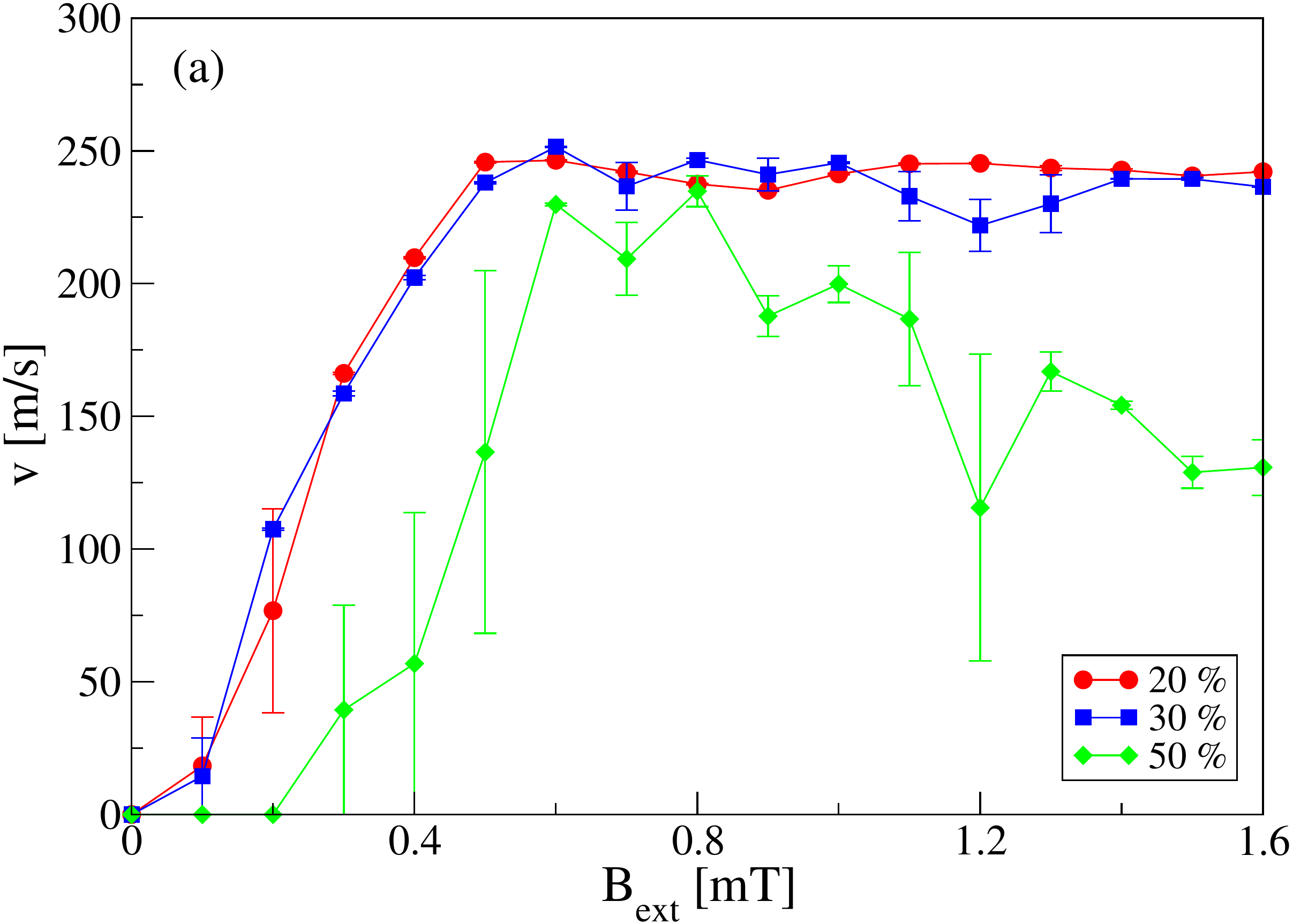}
\includegraphics[clip,width=0.48\textwidth]{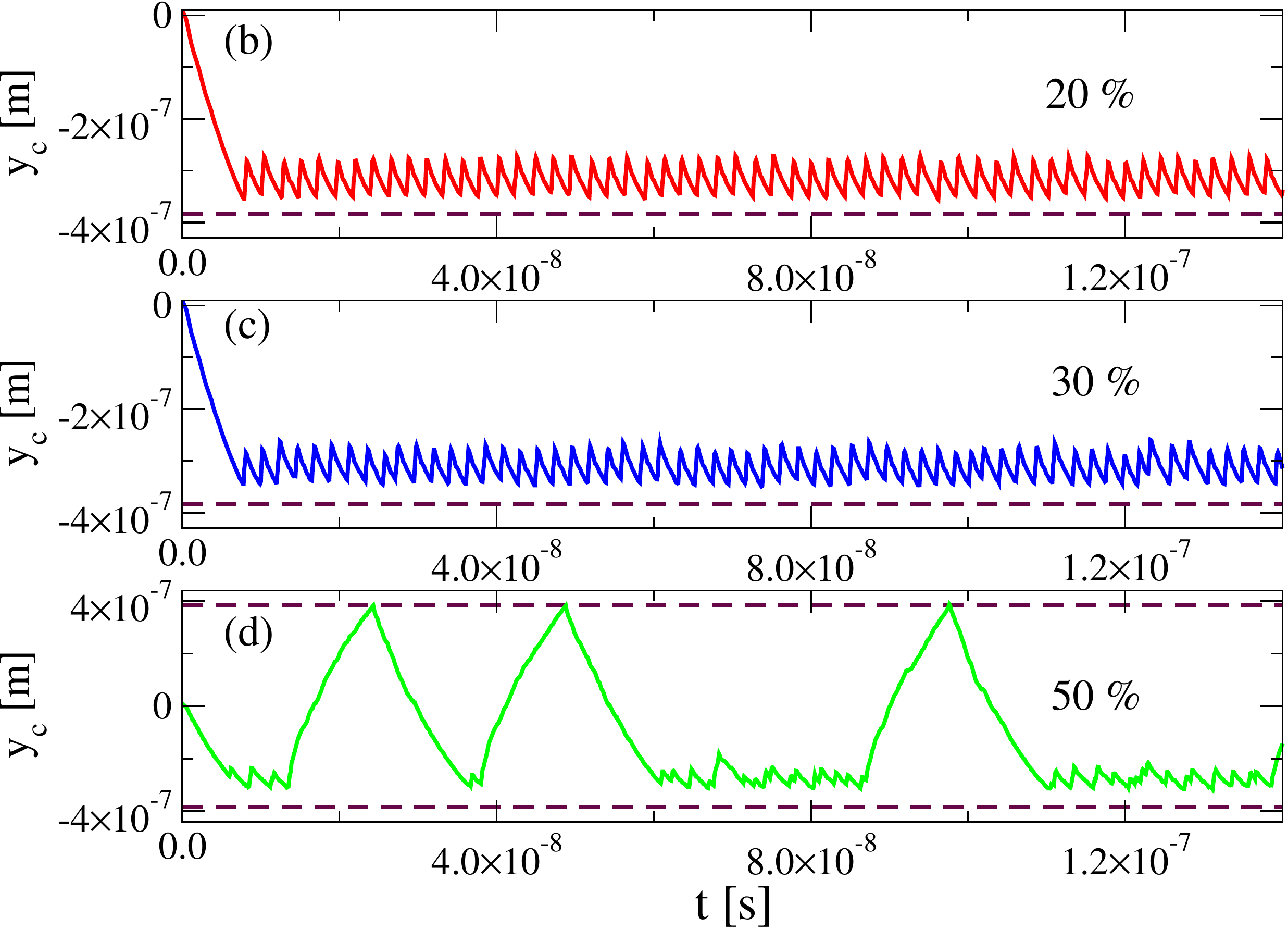}
\caption{(color online) (a) $v(B_\text{ext})$ curves for a strip with $w = 768$ nm and 
$\Delta_z = 15$ nm for different reductions of the exchange stiffness at the grain boundaries. 
(b)-(d) The trajectory $y_\text{c}$ of the vortex core for the strip shown in (a) at 
$B_\text{ext} =1.5$ mT for a reduction of the exchange stiffness of $20\%$, $30\%$ and $50\%$, 
respectively. The dashed lines represent the strip edges.}
\label{fig:Fig8_paperattrepul}
\end{figure}

\section{Quenched structural disorder}
\label{disorder}

Although systems with disorder are more relevant experimentally, the effect of disorder on 
DW dynamics has in relative terms (as compared to the ``pure'' systems) received less attention 
in recent numerical studies\cite{nakatani2003,min2010,leliaert2014prb,leliaert2014,vandewiele2012}. Here, we 
analyze the effect of quenched structural disorder on the DW dynamics in wide permalloy strips, 
focusing on the attraction-repulsion effect. To simulate the polycrystalline structure of the material, 
the system is divided in different areas by Voronoi tessellation. Each area represents a grain. We 
consider a grain size equal to the thickness of the strip, and study the effect of the reduction 
of exchange coupling between grains, by means of a reduction of the exchange stiffness $A_\text{exch}$ 
at the grain boundaries\cite{leliaert2014}. 

\begin{figure}[t!]
\leavevmode
\includegraphics[clip,width=0.48\textwidth]{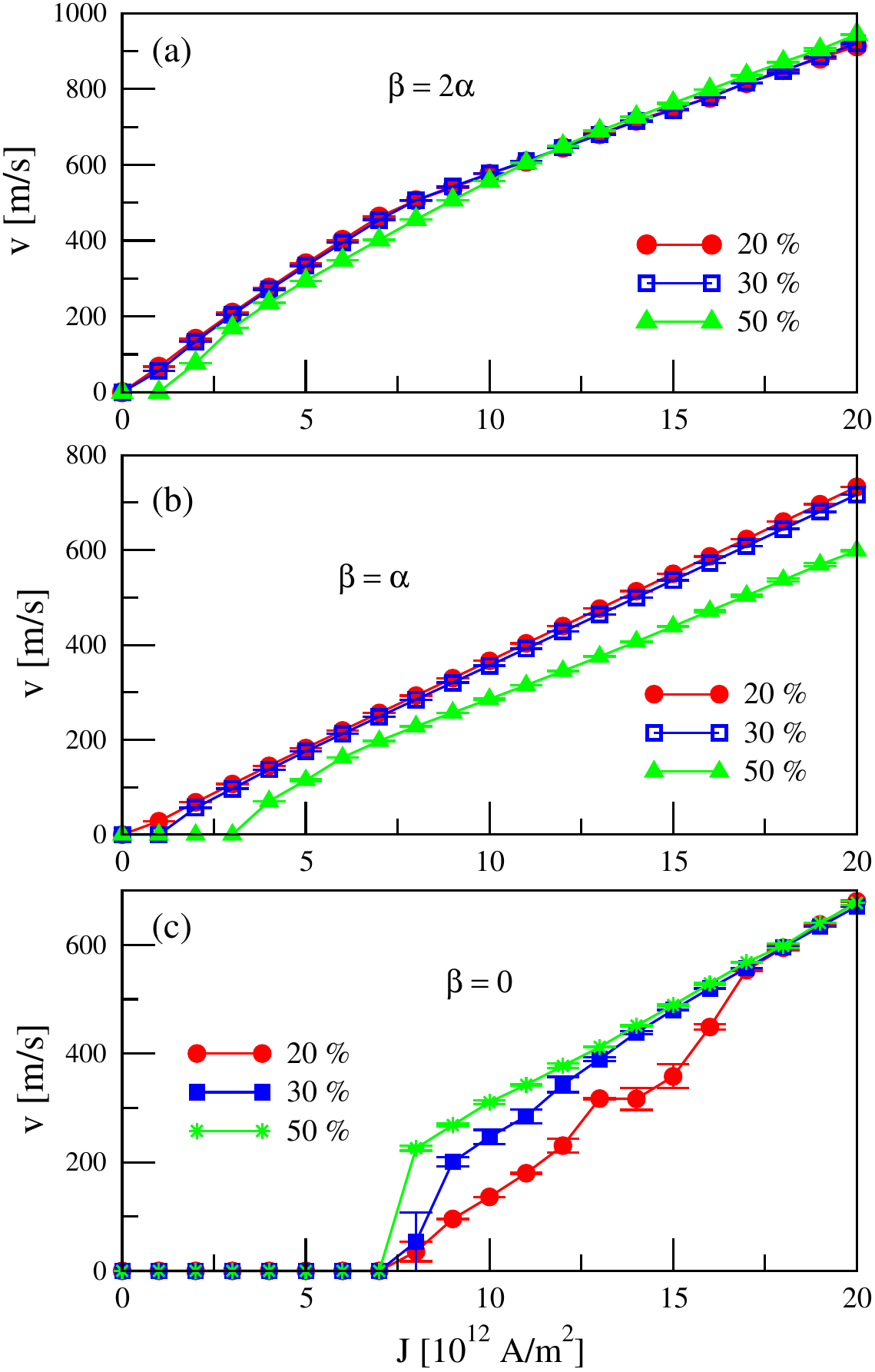}
\caption{(color online) $v(J)$ curves for a strip with $w = 768$ nm and $\Delta_z=15$ nm, considering 
several values of $\beta$, and three different values for the reduction of $A_\text{exch}$ at the grain 
boundaries. (a) $\beta = 2\alpha$. (b) $\beta = \alpha$. (c) $\beta = 0$.}
\label{fig:Fig9_paperattrepul}
\end{figure}

We start by considering the case in which the DW dynamics is driven by an external magnetic field. 
Contrary to the case of narrow strips where the vortex core often becomes pinned by disorder after some 
transient motion even for relatively large fields, for wide strips this is unusual. This is likely due to the
fact that in wide strips the vortex core can follow several paths. Fig.~\ref{fig:Fig8_paperattrepul} (a) 
shows the $v(B_\text{ext})$ curves for a strip with $w=768$ nm, $\Delta_z=15$ nm and different reductions of the 
exchange coupling across the grain boundaries. As the vortex motion depends on the random disorder 
configuration, the velocity represented is the average over $3$ different disorder realizations. For
disorder of moderate strength (with a reduction of exchange stiffness equal to $20\%$ or $30\%$), the 
velocity exhibits the characteristic plateau of the attraction-repulsion effect. The trajectories 
$y_\text{c}$ of the vortex core show oscillations close to the strip edge, see Figs.~\ref{fig:Fig8_paperattrepul} 
(b) and (c). As the disorder increases, the oscillations are more irregular. Thus, the attraction-repulsion
effect is robust against disorder of moderate strength. However, for strong enough disorder, e.g., with an exchange 
reduction of $50\%$, the vortex motion reflects a combination of transformations between different 
DW structures and oscillations close to the border. This can be clearly seen in Fig.~\ref{fig:Fig8_paperattrepul} 
(d), where the trajectory of the vortex core is shown. As a consequence of this combination, 
the average DW velocity decreases. On the other hand, for strong disorder the DW becomes pinned for 
low fields, see Fig.~\ref{fig:Fig8_paperattrepul} (a).

\begin{figure}[t!]
\leavevmode
\includegraphics[clip,width=0.48\textwidth]{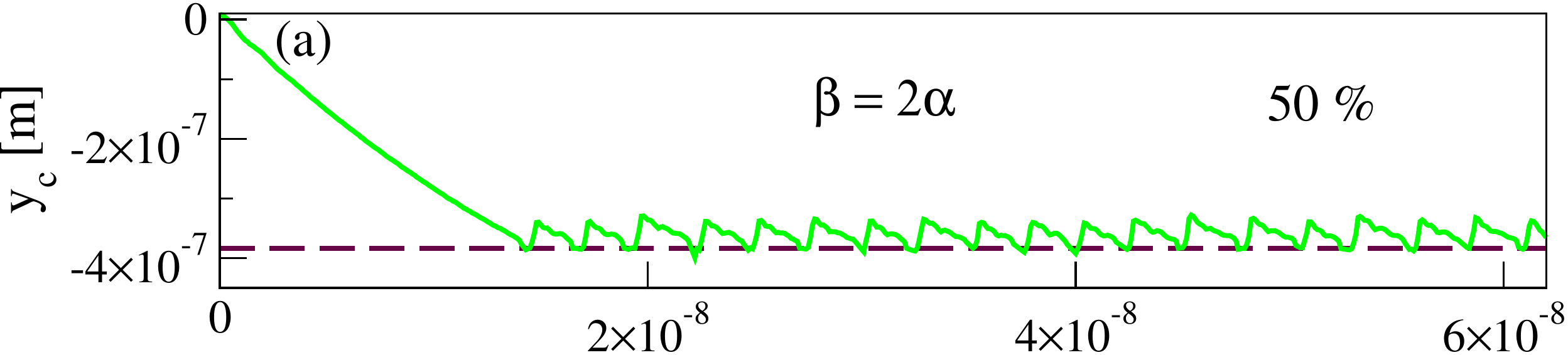}
\includegraphics[clip,width=0.48\textwidth]{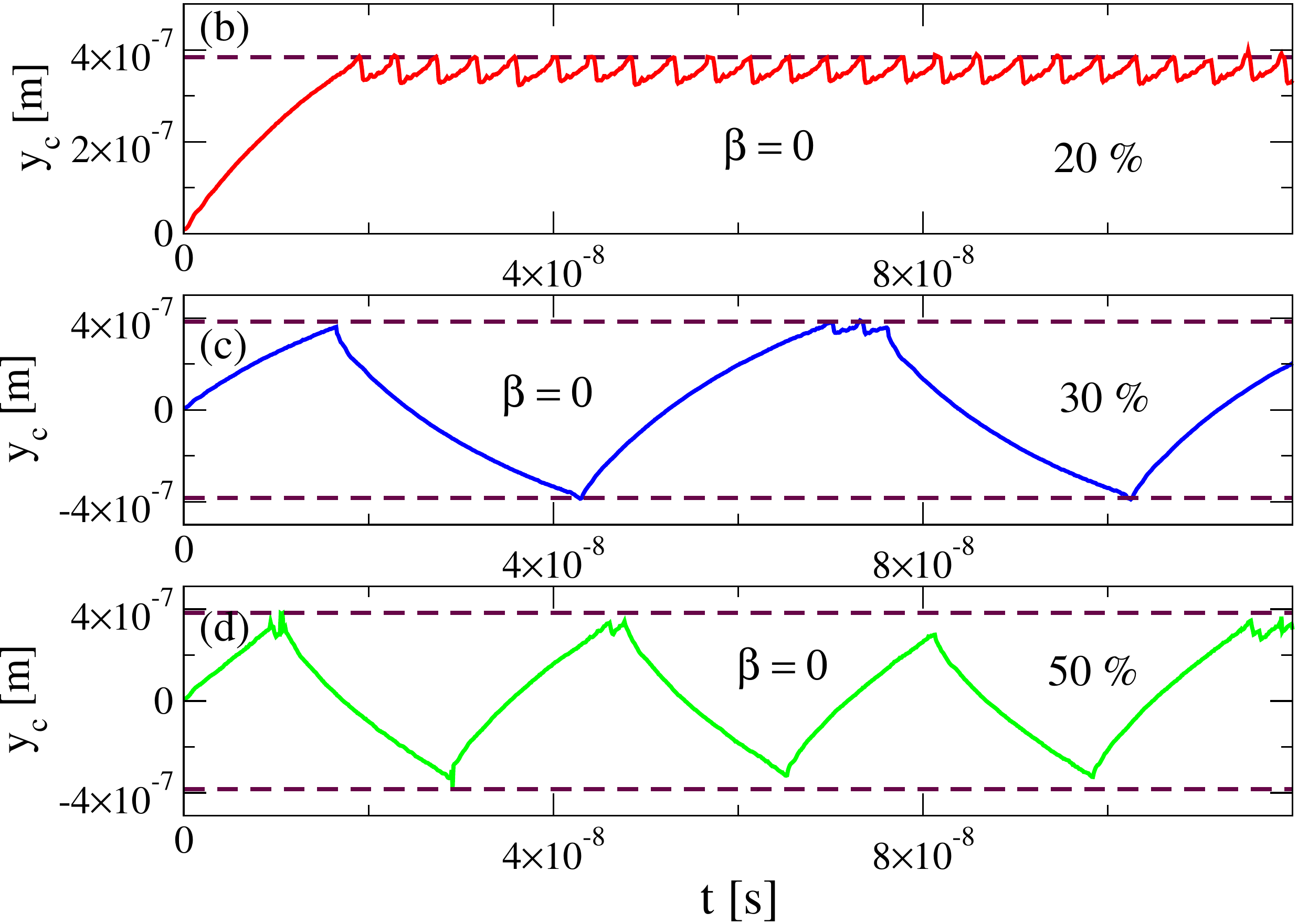}
\caption{(color online) Current-driven trajectories $y_\text{c}(t)$ of the vortex core for a strip with 
$w = 768$ nm and $\Delta_z = 15$ for different values of reduction of $A_\text{exch}$ across the grain 
boundaries, and different $\beta$-values. The dashed lines represent the strip edges. (a) $\beta= 2 \alpha$ 
and $J = 15\times 10^{12}$ A/m$^{2}$ and an exchange stiffness reduction of $50\%$, showing the 
attraction-repulsion effect. (b)-(d) Trajectories for the same strip represented in (a) in the case of 
$\beta=0$ and $J = 12\times 10^{12}$ A/m$^{2}$ for different reductions of $A_\text{exch}$.}
\label{fig:Fig10_paperattrepul}
\end{figure}

For the case of DW dynamics driven by spin-polarized currents, the effect of disorder depends strongly 
on $\beta$. For $\beta=2\alpha$, disorder barely affects the attraction-repulsion effect, see 
Fig.~\ref{fig:Fig9_paperattrepul} (a) where the $v(J)$ curves are represented
for a strip with $w=768$ nm, $\Delta_z=15$ nm and different reductions of the exchange coupling. As in 
the perfect case, there are two regimes of behavior, the steady state and the attraction-repulsion effect. 
The vortex core trajectory for this strip at  $J = 15\times 10^{12}$ A/m$^{2}$ and strong disorder is 
shown in Fig.~\ref{fig:Fig10_paperattrepul} (a). Even for strong disorder the oscillations of the core position
close to the strip edge are periodic. On the other hand, a finite ``extrinsic'' depinning threshold emerges
for strong disorder ($50\%$), see Fig.~\ref{fig:Fig9_paperattrepul} (a). The disorder affects slightly 
the slope of the velocity in the steady state. Moreover, the current density at which the regime 
with the attraction-repulsion effect starts is larger than in the case without disorder. For example, in 
a perfect strip with $J = 12\times 10^{12}$ A/m$^{2}$ the attraction-repulsion effect is observed, whereas
for a strip with strong disorder the dynamics is the typical one of the steady state. The disorder 
also softens the transition between the two regimes of behavior. At large values of $J$, $v$ is almost 
independent of the disorder.

Also for $\beta=\alpha$ and low values of $J$, a disorder-dependent depinning threshold is observed, see  
Fig.~\ref{fig:Fig9_paperattrepul} (b). The depinning threshold depends more strongly on the disorder strength
than in the $\beta=2\alpha$ case. Moreover, the velocity at a given $J$ decreases with the disorder
strength. For moderate disorder, the vortex dynamics is the same as that found in the perfect case. 
However, for strong disorder the vortex motion occurs with the vortex core displaced from the equilibrium 
position in the middle of the strip. This can be related to an effective damping parameter $\alpha^{\ast}$ 
generated by the disorder, as occurs for the adiabatic case in strips with nonmagnetic voids\cite{vandewiele2012}. 

For the adiabatic case ($\beta=0$) the behavior is completely different from the rest of the cases. Curiously, 
the velocity at a given $J$ increases with the disorder strength, see  Fig.~\ref{fig:Fig9_paperattrepul} (c). 
For a reduction of $20\%$, there are two regimes of behavior. For 
$8\times 10^{12} \text A/\text m^{2} \leq J \leq 14\times 10^{12}$ A/m$^{2}$ the vortex dynamics displays attraction-repulsion 
effect, while for $J \geq 15\times 10^{12}$ A/m$^{2}$ a combination of the attraction-repulsion effect and 
transformations between different DW structures is observed. As $J$ increases, transformations between the 
structures become the more prominent behavior. For $30\%$ and $50\%$, 
transformations between different DW structures are the main behavior. This can be clearly seen in 
Figs.~\ref{fig:Fig10_paperattrepul} (b)-(d), which show the vortex core trajectory for $J=12\times 10^{12}$ A/m$^{2}$ 
in a strip with $w=768$ nm, $\Delta_z=15$ nm and exchange stiffness reductions of $20\%$, $30\%$ and $50\%$, 
respectively. For moderate disorder, the oscillations close to the strip edge are periodic and the 
attraction-repulsion effect is stable. For stronger disorder, the attraction-repulsion effect almost 
disappears, leading to transformations between DW structures. As we already saw for perfect strips, in 
the adiabatic case the velocity is larger when the dynamics shows the transformations between DW structures than when the 
attraction-repulsion effect takes place. As the disorder increases, the transformations between the DW 
structures are more favored than the attraction-repulsion effect. For that reason, the velocity increases 
with the disorder. At high values of the spin-polarized current density, the velocity is independent of the 
magnitude of the structural disorder.

\section{Summary and conclusions}
\label{summary}

To summarize, we have studied in detail the attraction-repulsion effect, an oscillatory motion of 
the vortex core localized close to one of the strip edges. This behavior, characteristic of wide permalloy 
strips, avoids the transformations between different DW structures, leading to a plateau-like regime
with large values of the DW velocity, instead of the velocity drop observed in narrow strips. 
We have analyzed the attraction-repulsion effect when the DW dynamics is driven by
an external magnetic field, and also by spin-polarized electric currents. In both cases, we have 
seen that the origin of the oscillations is a repeated double switching of the magnetic vortex core. 
The first switching occurs via the creation of a vortex-antivortex pair, followed by an annihilation 
process. The second switching is assisted by the spin waves generated in the annihilation process. 
This double switching is a mechanism of the system to reduce its energy during the DW dynamics. 

We have also studied the effect of quenched structural disorder on the DW dynamics in wide permalloy 
strips. Depending on the driving force (field or current) the disorder affects the DW dynamics in 
different ways. For the case of a magnetic field, the attraction-repulsion effect is robust against 
quenched structural disorder of moderate strength. For DW dynamics driven by spin-polarized electric 
currents, the degree of non-adiabaticity $\beta$ plays an important role in the DW dynamics in 
disordered strips. For all values of $\beta$ the vortex motion can be pinned at very low $J$ due to 
the disorder. For $\beta = \alpha$, we have seen that the DW velocity at a fixed $J$ decreases with the 
disorder strength.  
For  $\beta = 2\alpha$, the attraction-repulsion effect is stable even in the case of strong disorder. 
However, in the adiabatic case ($\beta=0$) the attraction-repulsion effect only occurs for weak disorder. 
Curiously and contrary to what happens in other cases where the attraction-repulsion effect leads to large
values of the velocity, for $\beta=0$ the DW velocity is larger when there are transformations 
between different DW structures. As a result, DW velocity at a given $J$ increases with the disorder 
strength. Thus, for possible applications where stable and large values of the DW velocity as produced 
by the attraction-repulsion effect could be useful, the rather complex interplay of the details of the 
driving force and the disorder magnitude should be taken into account.

\begin{acknowledgments}
This work has been supported by the Academy of Finland through its Centres
of Excellence Programme (2012-2017) under project no. 251748, and an Academy
Research Fellowship (LL, project no. 268302). We acknowledge the computational 
resources provided by the Aalto University School of Science ``Science-IT'' 
project, as well as those provided by CSC (Finland).
\end{acknowledgments}

\end{document}